\documentclass[aps,prl,reprint, amsmath,amssymb,superscriptaddress]{revtex4-1}
\usepackage{graphicx}

\begin{document}

\title{Primary and Secondary Order Parameters in the Fully Frustrated Transverse Field Ising Model on the Square Lattice}

\author{Gabe Schumm}
\email{gschumm@bu.edu}
\affiliation{Department of Physics, Boston University, 590 Commonwealth Avenue, Boston, Massachusetts 02215, USA}

\author{Hui Shao}
\affiliation{Center for Advanced Quantum Studies, Department of Physics, Beijing Normal University, Beijing 100875, China}
\affiliation{Key Laboratory of Multiscale Spin Physics, Ministry of Education, Beijing 100875, China}

\author{Wenan Guo}
\affiliation{Department of Physics, Beijing Normal University, Beijing 100875, China}
\affiliation{Key Laboratory of Multiscale Spin Physics, Ministry of Education, Beijing 100875, China}

\author{Fr\'ed\'eric  Mila} 
\email{frederic.mila@epfl.ch}
\affiliation{Institute of Physics, Ecole Polytechnique F\'ed\'erale de Lausanne (EPFL), CH-1015 Lausanne, Switzerland}

\author{Anders W. Sandvik}
\email{sandvik@bu.edu}
\affiliation{Department of Physics, Boston University, 590 Commonwealth Avenue, Boston, Massachusetts 02215, USA}
\affiliation{Beijing National Laboratory for Condensed Matter Physics and Institute of Physics, Chinese Academy of Sciences, Beijing, 100190, China}

\date{\today}

\begin{abstract}
 Using quantum Monte Carlo simulations and field-theory arguments, we study the fully frustrated
transverse-field Ising model on the square lattice for the purpose of quantitatively relating two
different order parameters to each other. We consider a ``primary'' spin order parameter and a
``secondary'' dimer order parameter, which both lead to the same phase diagram but detect $\mathbb{Z}_8$ and $\mathbb{Z}_4$ symmetry breaking, respectively. While at $T>0$ their scaling exponents are simply related to each other,
as explained by a mapping to a height model, we show that at $T=0$ they correspond to different charge sectors 
of the O(2) model in 2+1 dimensions with non trivial exponents that are not simply related to each other.
Our insights are transferrable to a broad class of Ising models whose low-energy physics involves dimer 
degrees of freedom, and also serve as a guide to treating primary and secondary order parameters more generally.

\end{abstract}

\maketitle

The concept of an order parameter is key to quantitative descriptions of phase transitions. In some systems it is natural to define more than 
one order parameter, either in some trivial way or using emergent degrees of freedom originating from some mapping to an effective 
low-energy model. The relationships between different order parameters may be non-trivial, e.g., unexplained behavior of a 
``parasitic'' ferromagnetic order parameter in a system with primarily antiferromagnetic order was reported \cite{hattori_14, hattori_16, tsunetsugu_21}. Here we consider a quantum spin model that very clearly illustrates two different 
order parameters that not only exhibit different scaling behaviors but the relationships between the critical exponents of the order parameters
are also different at temperature $T=0$ and $T>0$. 

\begin{figure}[b]
\centering
\includegraphics[width=77mm]{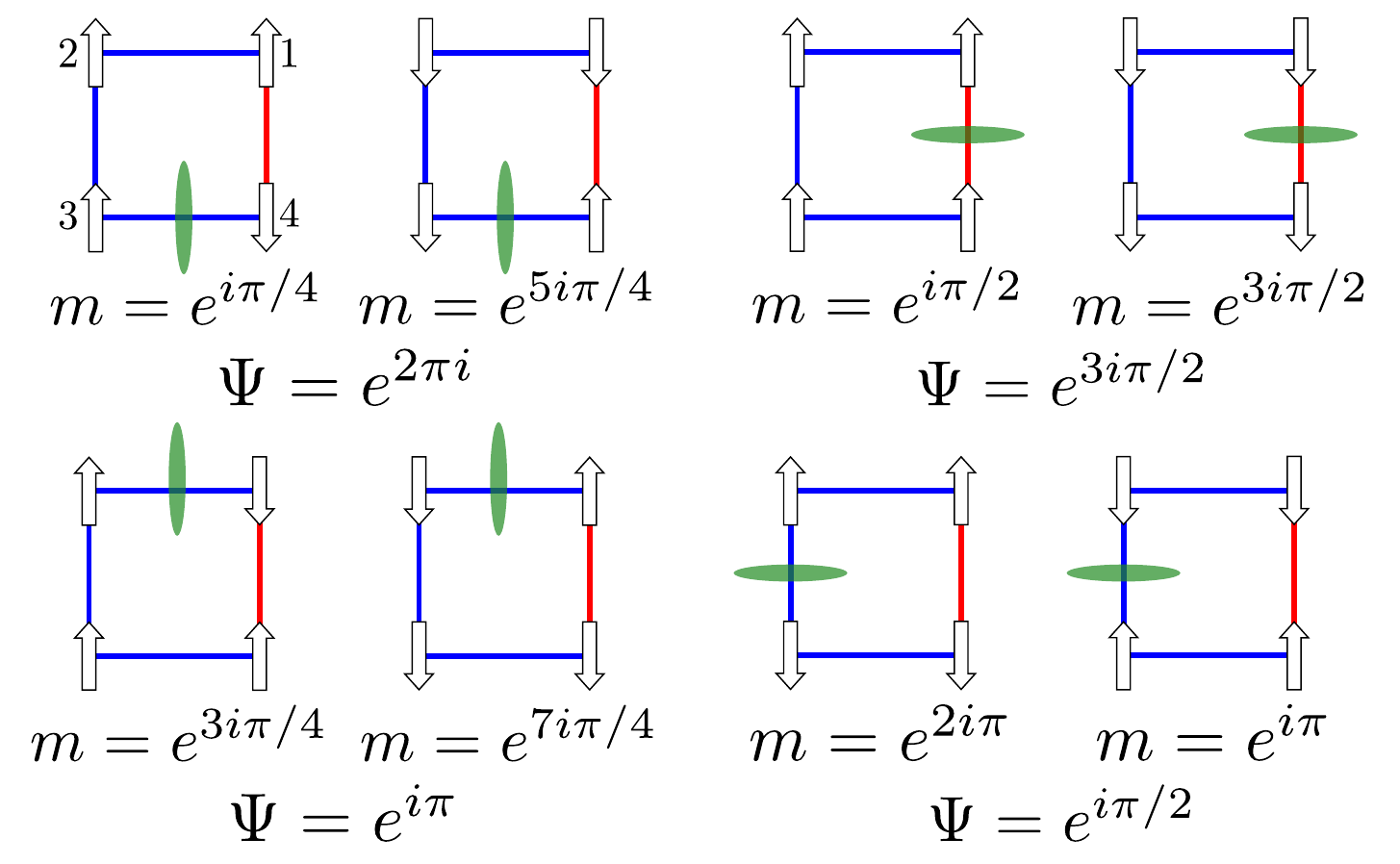}
\caption{The degenerate ground states represented by elementary plaquettes. Blue and red bonds show $J_{ij} = -J$ and $J_{ij} = +J$,
respectively. The direction of the sublattice magnetizations $m_s$, $s=1,\ldots,4$, are indicated by the arrows, and dimers (green ovals) are defined on the
frustrated bonds. The values of the magnetization and dimer order parameters shown correspond to the ground state sublattice magnetizations of the stacked, classical ($\Gamma = 0$) model \cite{blankschtein_84_sq},  $\sin(\pi/8)$ for the sites sharing a frustrated bond and $\cos(\pi/8)$ for the two others [see Eqs. (3) and (5)].} 
\label{col}
\end{figure}

We study the two-dimensional (2D) square-lattice fully frustrated transverse field (Villain) quantum Ising model (FFTFIM), with Hamiltonian
\begin{equation}\label{Ham}
  H = \sum_{\langle{ij}\rangle} J_{ij} \sigma_i^z\sigma_j^z - \Gamma \sum_j \sigma^x_j,
\end{equation}
where $\sigma_i^x$ and $\sigma_i^z$ are Pauli operators. The couplings $J_{ij}$ are equal in magnitude but the number of antiferromagnetic  (AF) 
couplings around any elementary plaquette is odd \cite{villain}, here with $J_{ij} = +J$ (AF) on every second column and 
$J_{ij} = -J$ on all other bonds as depicted in Fig.~\ref{col}. The classical model at $\Gamma=0$ hosts a large ground state degeneracy that is 
lifted by the transverse field via an ``order-by-disorder'' mechanism \cite{villain_order_disorder, moessner_order_disorder, henley_order_disorder}. 

The first studies of the FFTFIM considered the stacked version of the classical  model using a using Landau-Ginsburg-Wilson 
(LGW) approach \cite{blankschtein_84_sq} as well as Monte Carlo (MC) simulations \cite{sachdev_91}. The LGW study predicted an eight-fold degenerate ground 
state, which was more precisely characterized by Ref.~\cite{sachdev_91} as a $\mathbb{Z}_4$ symmetry breaking phase, corresponding to $90^{\circ}$ 
rotations of the lattice, paired with a global spin-flip symmetry.

The model was later treated using quantum MC (QMC) simulations \cite{wenzel_mila}, where spin and dimer order was found at $T=0$ for 
$\Gamma < \Gamma_c \sim 1.578$. In this phase the frustrated bonds (mapped to dimers as in Fig.~\ref{col}) align along alternating columns 
or rows; the $\mathbb{Z}_4$ symmetry breaking phase identified in the stacked model. The order parameters considered previously detected 
$\mathbb{Z}_4$ symmetry and spin-reflection symmetry separately. The reconciliation of $\mathbb{Z}_8$ versus $\mathbb{Z}_4$ breaking 
was touched on by Coletta et al.~\cite{coletta}, but the relationship between the respective order parameters was not explored.

Here we define a proper $\mathbb{Z}_8$ symmetric spin order parameter and demonstrate that the $\mathbb{Z}_4$ dimer order parameter should be considered as
\textit{secondary}. While both order parameters lead to the same phase diagram (provided in Supplemental Material, Sec.~SI \cite{supp}), the $\mathbb{Z}_4$
order parameter exhibits faster decaying critical correlations. Both order parameters, when correctly defined, exhibit emergent U(1)
symmetry in the critical phase as well as at the quantum phase transition, stemming from the irrelevance of the discrete symmetry-breaking terms at criticality \cite{jose}. However, the distinction between primary and secondary order is made quantitative by considering the critical scaling in these two regimes. This is only made possible by connecting these order parameters to operators in the relevant field theories,
at $T=0$ and $T>0$. Finite-size scaling of QMC (stochastic series expansion \cite{sandvik03}) 
results of the full FFTFIM Hamiltonian, Eq.~(\ref{Ham}), support the predicted scaling, emphasizing the utility of this novel approach to studying quantum magnetism. 

Secondary order parameters have  previously been used to describe higher harmonic contributions to spatial modulation in density wave systems,
e.g., liquid crystals \cite{wu_94,aharony_95,OPE_O2_3,netz_97,calabrese_05}. A secondary order parameter can clearly be defined also in the
FFTFIM, but the different scaling forms of the spin and dimer order parameter in the FFTFIM have not been addressed. This Letter provides a framework 
for secondary order not just in the FFTFIM, but in the entire class of frustrated Ising models with effective dimer degrees of freedom, e.g., 
the antiferromagnet on the triangular lattice \cite{blankschtein_84_tri,blote_93, isakov, moessner_sondhi_01,stephenson_70}.

\textit{Order Parameters.}---To construct a proper primary order parameter, we follow standard procedures \cite{blankschtein_84_sq,isakov,coletta}, 
using an effective Hamiltonian for the amplitude $m$ and phase $\theta$ of critical modes:
\begin{equation} \label{LGW}
\begin{split}
H_{\rm eff} = \sum_{\vec{q}} (r + q^2)|m|^2 + u_4 |m|^4 + u_6 |m|^6 +\\ (u_8 + v_8/32)|m|^8 - (v_8/32)|m|^8 \cos(8\theta).
\end{split}
\end{equation}
The eight-state clock anisotropy implies an eight-fold degenerate ground state, characterized by sublattice magnetizations
$(m_1, m_2, m_3, m_4)$; see Fig. \ref{col}. Each state corresponds to one frustrated bond in a plaquette, where the magnitude of the sublattice magnetizations of the sites sharing this bond are smaller than the other two, $\sin(\pi/8)$ and $\cos(\pi/8)$ respectively \cite{blankschtein_84_sq}. An overall spin-flip
transformation gives a total of eight degenerate states.  

Based on the low-energy behavior of the stacked model, as well as the semiclassical analysis of Ref.~\cite{coletta}, we define the primary
order parameter as the complex number 
\begin{equation}\label{mag_eq}
\begin{split}
m = &\,m_x + im_y =\\ & \frac{1}{2} \left(m_1e^{i\frac{\pi}{8}} + m_2e^{i\frac{3\pi}{8}} + m_3e^{i\frac{5\pi}{8}} + m_4e^{i\frac{7\pi}{8}}\right)
\end{split}
\end{equation}
where 
\begin{equation}
m_s = \frac{4}{N} \sum_{j \in s} \sigma_j^z.
\end{equation}
The eight ground state configurations of the stacked model correspond to $m=e^{i n\pi/4}$, $n = 1, 2, \dots, 8$.

The problem can also be mapped onto that of dimer coverings, where a dimer is assigned across each frustrated bond \cite{moessner_sondhi_01}, and
we define a secondary order parameter $\Psi$ with this mapping in mind. This order parameter is also complex number, defined in terms of the dimer
density modulation on the dual lattice \cite{wenzel_mila}:
\begin{equation}\label{dim_eq}
    \Psi = \Psi_x + i\Psi_y = 2\tilde{d}_x(0, \pi) +  2i\tilde{d}_y(\pi, 0),
\end{equation}
where 
\begin{equation}
  \tilde{d}_\alpha(\boldsymbol{q}) = \frac{1}{N} \sum_i e^{i\boldsymbol{q}\cdot \boldsymbol{r}_{i}} d_{i, \alpha},
\end{equation}
is the Fourier transformed dimer density
\begin{equation}\label{dimer}
  d_{i, \alpha} = 1 + \frac{J_{i, j_\alpha}}{J}\sigma_i^z \sigma_{j_\alpha}^z,
\end{equation}
and $j_\alpha$ is the index of nearest neighbor to site $i$ in the $\alpha$ direction. Long-range ordering is associated with $|\Psi|$ taking a 
finite value, while the specific (symmetry-broken) ordering pattern is identified by the phase. Thus, $\Psi$ takes one out of the values 
$e^{i n\pi/2}$, $n = 1, 2, \dots, 4$. The connection between the columnar states and the sublattice magnetizations is illustrated in Fig.~\ref{col}. 

\textit{$\mathbb{Z}_8$ versus $\mathbb{Z}_4$ Symmetry Breaking.}---We detect the order-parameter symmetries by plotting the probability distributions of $m$ and
$\Psi$ accumulated during QMC simulations. In the ordered phase, we expect eight (four) $\delta$-functions at the eight (four) values corresponding
to the columnar spin (dimer) states in Fig.~\ref{col}. These $\delta$-functions smear for finite systems, appearing as highly peaked Gaussian distributions
for large systems.
 
In the critical phase, we observe the emergent U(1) symmetry expected from the mapping to the height model \cite{emig,korshunov,henley_97}.
We assign height differences to neighboring spins based whether or not the bond they share is frustrated (i.e., crossing a dimer) \cite{korshunov},
as detailed in Supplemental Material, Sec. SII \cite{supp}. In the ordered phase, the height profile is ``flat,'' with the height values
bounded from above. In the critical and disordered phases, the model is in its ``rough'' phase, with a logarithmically diverging height profile.
This behavior can be described  by an effective elastic free energy with a periodic ``locking'' potential favoring the eight flat height
configurations, i.e., the columnar states in Fig. \ref{col}. This effective free energy is precisely that of the 2D $XY$ model, where the
locking potential corresponds to an $q=8$ state clock anisotropy term. This connection allows us to apply the renormalization group (RG)
analysis of Ref.~\cite{jose} to understand the observed behavior.

Jose et al.~\cite{jose} first showed that the classical 2D $q$-state clock model is characterized by three temperature regimes if $q > 4$.
At temperatures below the lower critical temperature $T_{\rm c1}$, the clock term is relevant and the system orders into the $\mathbb{Z}_q$ clock phase.
At $T$ above $T_{\rm c1}$ but below the upper critical temperature $T_{\rm c2}$, the clock term is irrelevant and the free energy reduces to that
of the $XY$ model in the KT phase. In this phase, the system can freely fluctuate between the flat height configurations, thus resulting in the U(1)
symmetric distributions that we observe. Finally, above $T_{\rm c2}$ the critical phase melts into the disordered phase as defects proliferate.

\begin{figure}[t]
\includegraphics[width=75mm]{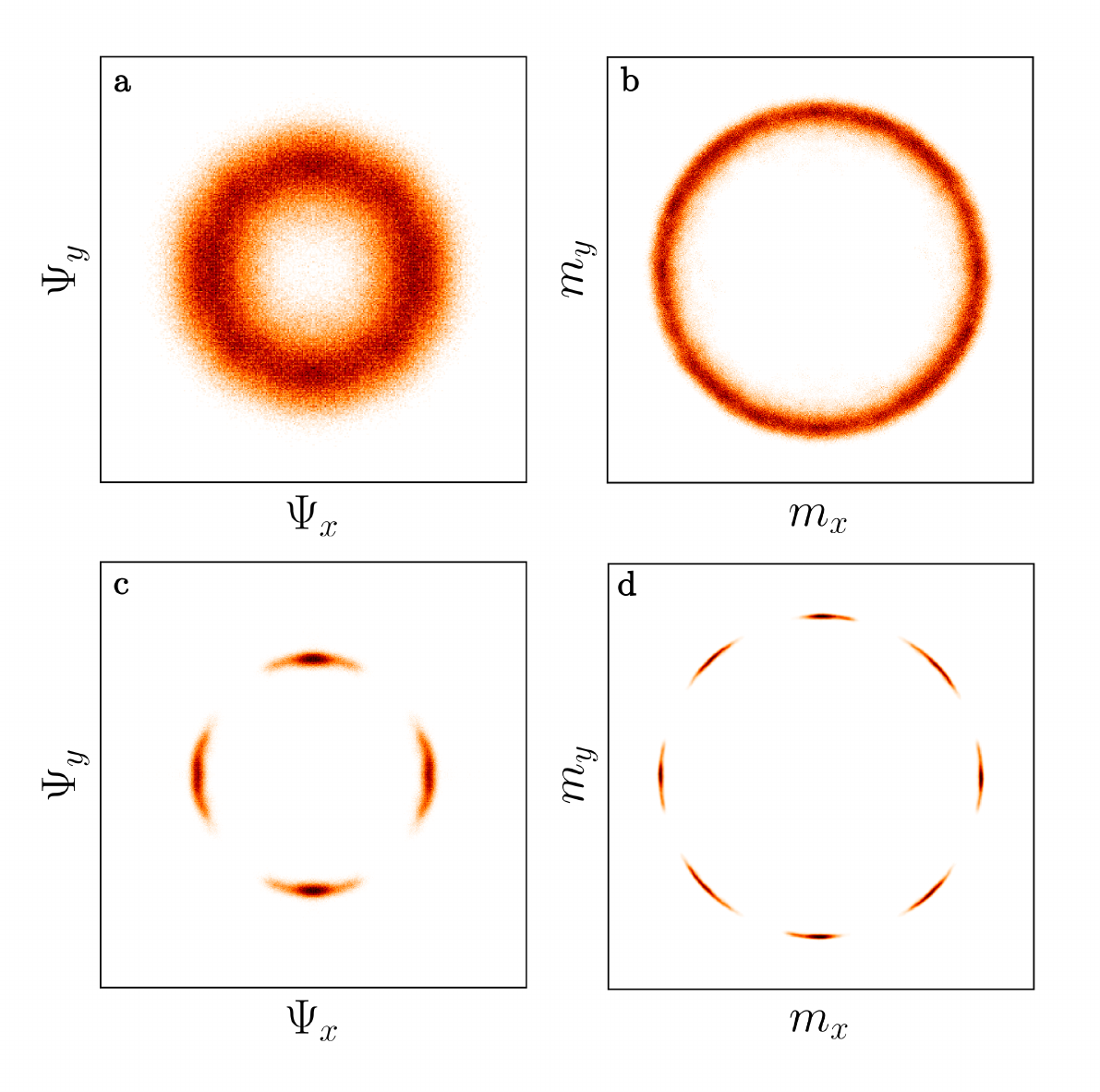}
\caption{Distributions of $(\Psi_x, \Psi_y)$ in (a,c) and $(m_x, m_y)$ in (b,d), collected in several independent simulations (each
  initialized in one of the four columnar states, to prevent trapping in states with topological defects) and symmetrized using lattice rotations
  and reflections. The system size is $L=96$ and $\Gamma/J = 0.43$, with $T/J = 0.21$ in (a,b) (in the critical phase)
  and $T/J = 0.014$ in (c,d) (in the ordered phase).}
\label{hist}
\end{figure}

An example of the symmetry reduction in the ordered phase is shown in Fig.~\ref{hist}, where (a) and (b) are collected from simulations in the
critical phase, while (c) and (d) are from the ordered phase. While we detect emergent U(1) symmetry in both order parameters, a U(1) phase
would not be expected for a primary $\mathbb{Z}_4$ dimer order parameter at $T>0$, given the $q>4$ criterion in the clock model \cite{jose}. 
However, with the spin order parameter corresponding to $q=8$, the critical phase is expected.

\begin{figure}[t]
\begin{center}
\includegraphics[width=\columnwidth]{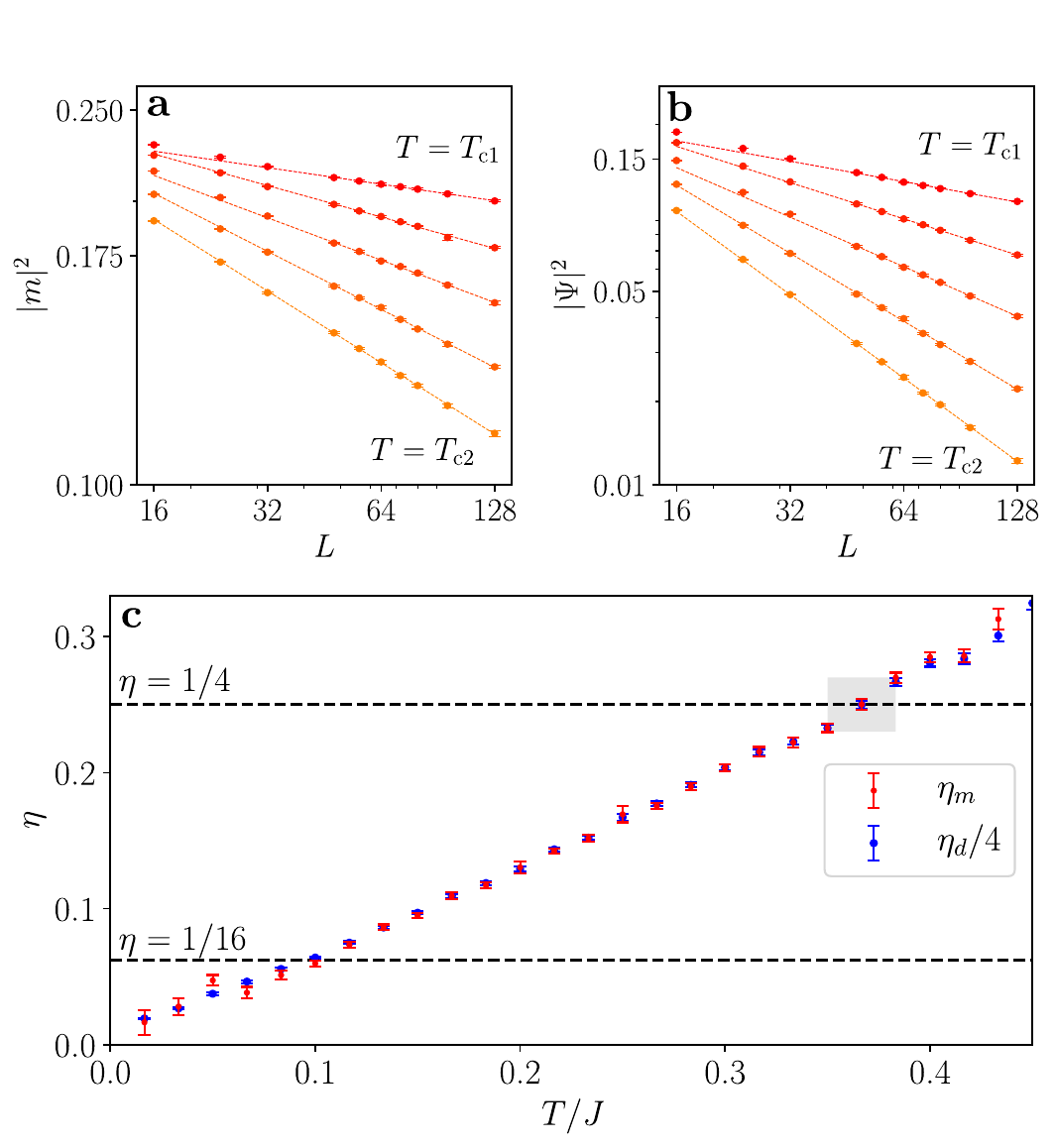}
\caption{Log-log plots of the magnitude of the primary (a) and secondary (b) order parameter versus the system size $L$ for a range of
temperatures at $\Gamma/J = 0.67$. The dashed lines are power-law fits to the largest eight system sizes, with $T$ increasing with color
brightness (red to orange). (c) Anomalous dimensions $\eta_{m,d}$ versus $T$ for the primary (red) and secondary (blue) order parameters,
extracted from fitting data to Eq.~(\ref{power_law}). The exponents align if $\eta_d$ is rescaled by a factor $1/4$. The gray box denotes
$T_{\rm c2} \pm \sigma$ from Binder crossing results. The dashed lines are at the predicted
values at the phase boundaries.}
\label{eta}
\end{center}
\end{figure}

\textit{Scaling at $T>0$.}---As a quantitative characterization of the primary and secondary natures of the
two order parameters, we compare the scaling of their respective correlation functions in the critical phase. Within the $q$-state clock-model
description, the spin-spin correlations should decay algebraically with a scaling exponent $\eta$ that varies continuously with
the temperature \cite{blote_84}. The value of $\eta$ at the upper and lower critical temperatures are known, $\eta=1/4$ and $\eta=4/q^2$,
respectively \cite{jose,henley_97}. To extract $\eta$, we examine the magnitudes of both order parameters, which in this phase
should scale with the lattice length $L$ as 
\begin{equation} \label{power_law}
  |m|^2 \propto L^{-\eta_m},~~~~|\Psi|^2 \propto L^{-\eta_d},
\end{equation}
where we leave open the possibility that $\eta_m \neq \eta_d$. In Fig.~\ref{eta}(a) and \ref{eta}(b), the order parameters are plotted versus
system size for a range of temperatures between $T_{\rm c1}$ and $T_{\rm c2}$. As predicted, they scale algebraically with $L$, and we extract $\eta_m(T)$
and $\eta_d(T)$ by fitting data to Eq.~(\ref{power_law}). Results are shown versus $T$ in Fig.~\ref{eta}(c).

The primary order parameter scales with the expected exponent $\eta_m=1/4$ at the upper critical temperature $T_{\rm c2}$ extracted from Binder
crossing results  (Supplemental Information, Sec.~SI \cite{supp}). Below this temperature, the exponent linearly decreases. The statistical quality
of the fits deteriorates below the temperature at which $\eta_m=1/16$ (see Supplemental Material, Sec.~SIII \cite{supp}), which is the predicted
value at the lower transition point $T_{\rm c1}$ \cite{jose}, where the system orders. While the linear decrease in $\eta$ appears to continue below $T_{\rm c1}$, the deterioration of the power-law fit used to extract the exponents below this temperature implies that this trend should not be given any credence. Thus, our numerical results are consistent with the theory, and
we can use $\eta_m=1/4$ and $\eta_m=1/16$ to set more precise upper and lower boundaries. 

The same behavior is observed for the dimer order
parameter, except that the value of the $\eta_d$ is consistently approximately four times larger than $\eta_m$. After rescaling $\eta_d$ by a
factor of four, the results match within statistical errors, suggesting the relation $\eta_d = 4\eta_m$. 
This relationship between $\eta_m$ and $\eta_d$ can be explained by the height model: the dimer value $d(\vec{r})$ at $\vec{r}$,
is related to the product of neighboring spin operators, Eq.~(\ref{dimer}), which in the coarse-grained model can be considered simply as the
square of the spin operator at $\vec{r}$, $\left(\sigma^z(\vec{r})\right)^2$. By representing the spin operators in terms of the height
variables, one can relate the spin-spin and dimer-dimer correlation functions to the logarithmically diverging height difference profile; the
observed factor of four relating $\eta_d$ and $\eta_m$ then emerges. For details, see Supplemental Material, Sec.~SII \cite{supp}.

\textit{Scaling at $T=0$.}---At the quantum critical point, there must be a different relationship between $\eta_d$ and $\eta_m$.
Given the irrelevance in 2+1 dimensions of $\mathbb{Z}_4$ or $\mathbb{Z}_8$ perturbations to a U(1) order parameter, the primary order parameter should scale with the
conventional 3D $XY$ critical exponent $1 + \eta_{\mathrm{3D}XY}$ \cite{3D_XY}, as previously confirmed in simulations with a spin-based order
parameter \cite{wenzel_mila}. However, to explain the critical scaling of the dimer order parameter, we must reference other aspects of the field theory.
Here we analyze both order parameters using simulations at $T=1/L$.

We extract $\eta_m$ from the asymptotic long distance ($r = L/2$) critical spin-spin correlation function, which is expected to scale as
$C_M(L/2)  \sim {L^{-(1+\eta_m)}} = {L^{-2\Delta_\phi}}$, where $\Delta_\phi$ is the scaling dimension of the operator
of the order parameter $\phi$ in the 3D O(2) theory. This is often referred to as a charge-1 (or spin-1) operator \cite{OPE_O2_1,OPE_O2_3},
indicating that it corresponds to a perturbation, e.g., $h\cos(\theta)$, inducing order in a single direction in the O(2) space, so that the
degeneracy is completely lifted. A corresponding perturbation in FFTFIM would be one that fully breaks the $\mathbb{Z}_8$ symmetry in the ordered
state, favoring one of the eight columnar spin configurations.

\begin{figure}[t]
\begin{center}
\includegraphics[width=80mm]{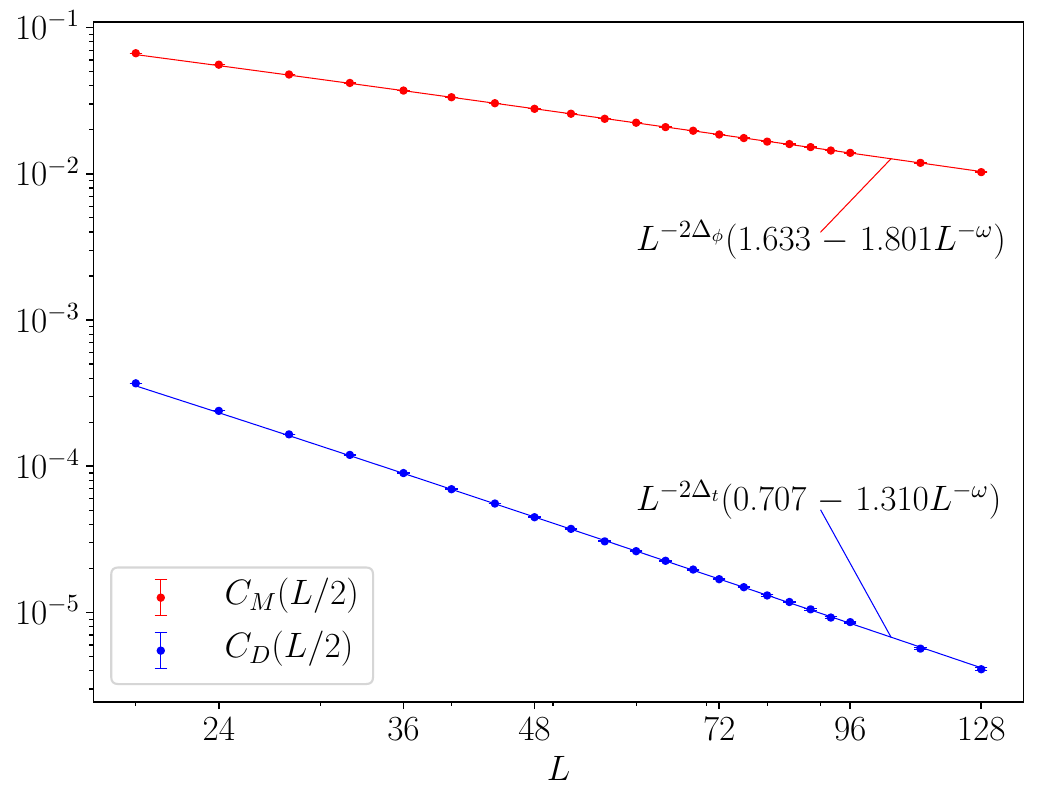}
\vskip-3mm
\caption{Dimer-dimer (blue) and spin-spin (red) correlation functions at the quantum critical point, $\Gamma/J = 1.5768$. The lines are fits of the 
$L\ge 44$ data to the expected scaling form
$\propto L^{-2\Delta_{s,t}}(a+bL^{-\omega})$, with $\Delta_\phi= 0.519088$, $\Delta_t= 1.23629$ \cite{OPE_O2_1}, and the correction exponent $\omega = 0.789$ \cite{hasenbusch_19}
in the 3D O(2) universality class. The reduced $\chi^2$ values are 
$\sim 0.8$ and $\sim 0.9$ for the spin and dimer correlations respectively.}
\label{scaling_dim}
\end{center}
\end{figure}

A perturbation that couples an external field to the secondary order parameter would not fully break the $\mathbb{Z}_8$ symmetry of the ground
state, but would favor the two spin configurations of given columnar dimer state. Accordingly, in the low-energy U(1) theory the perturbation
should be charge-2 (or spin-2 traceless symmetric) of the form $h\cos(2\theta)$, which can also be accomplished with products of $\phi$
components, e.g., $h\phi_1\phi_2$. This operator has scaling dimension often referred to as $\Delta_t$ \cite{OPE_O2_1, OPE_O2_3},

To test the scaling form ${L^{-(1+\eta_d)}}  = {L^{-2\Delta_t}}$, we analyze the oscillating part $C_D(L/2)$ of the dimer-dimer correlation function.
Since the connection between the primary order parameter and $\Delta_\phi$ is well known, we first used its scaling behavior to refine the value of 
$\Gamma_c$ reported in Ref.~\cite{wenzel_mila}, as detailed in Supplemental Material, Sec.~SV \cite{supp}, obtaining $\Gamma_c=1.57680 \pm 0.00009$. 
We then calculated $C_M(L/2)$ and $C_D(L/2)$ at the midpoint; their scaling behaviors are shown in
Fig.~\ref{scaling_dim}. The results match very well the scaling dimensions obtained in recent numerical conformal bootstrap calculations \cite{OPE_O2_1}:
$\Delta_\phi \approx 0.519088$ and $\Delta_t \approx 1.23629$.

The simplest effective model with the same microscopic symmetries and exhibiting the same scaling behavior is a classical 3D 8-state clock model,
where a charge-$l$ order parameter is defined by the vector $\vec{m}_l = (m_x,m_y)$, with $m_x=\sum_i \cos(l\theta_i)$, $m_y=\sum_i \sin(l\theta)$, 
$\theta_i$ being the angle of spin $i$. Results for this model are presented in Supplemental Material, Sec SVI \cite{supp}). The excellent
agreement with the expected exponents in both the FFTFIM and clock model confirms without doubt the emergent U(1) symmetry and the primary 
and secondary nature of the order parameters in the FFTFIM.

\textit{Conclusion.}---We have clarified the nature of the two order parameters, based on spins and dimers, in the square-lattice FFTFIM.
The spin-based order parameter is primary, as it scales with the leading critical exponents and displays the full eight-fold degeneracy
of the ordered phase, while the $\mathbb{Z}_4$ dimer order parameter is secondary with faster decaying critical correlations. Our QMC results at
$T=0$ and $T>0$ confirm the scaling exponents predicted from the respective low-energy field theories.

Mapping to dimer models is a powerful tool in the study of frustrated spin systems, and the problem of lattice coverings by hard-core
dimers is an interesting topic in itself. The connection between the secondary dimer order parameter and the primary spin order parameter
then provides a crucial link between the models that had not been previously drawn in this context. Beyond the particular FFTFIM
considered here, the triangular and kagome lattice AF Ising models \cite{nagai_93, moessner_sondhi_01}, the fully-frustrated honeycomb lattice Ising model
\cite{moessner_sondhi_01}, and the fully frustrated 4-8 lattice Ising models \cite{hearth_22} are all well studied systems where the secondary
order parameter prescription could also be applied. 

Our insights also explain the critical scaling of a so-called ``parasitic'' order parameter studied in the AF 3-state Potts model on the 
diamond lattice \cite{hattori_14, hattori_16, tsunetsugu_21}. While this system primarily orders antiferromagnetically in the ground state, it was shown that the presence of this order induces a finite ferromagnetic moment, which is captured by a secondary order parameter. The observed scaling of this order parameter had previously eluded explanation, but it is now clear that this ``parasitic''
order parameter also has scaling dimension $\Delta_t$ in that system.

The relevance of the secondary order parameter operator ($\Delta_t < 3$) in the FFTFIM implies that a perturbation favoring
one of the dimer states, accomplished by appropriately modulating the Ising couplings, will induce a $\mathbb{Z}_2$ symmetry breaking
phase in the plane of $\Gamma$ and the dimer field (modulation) strength $h_d$. The phase boundary between the paramagnetic and ordered phases 
should have the asymptotic form $h_{d,c} \sim |\Gamma_c(h_{d,c}) - \Gamma_c(0)|^{\nu/\nu_d}$, where $\nu$ is the 3D O(2) correlation-length 
exponent and $\nu_d=(3-\Delta_t)^{-1}$, which we confirm in Supplemental Material, Sec.~SVII \cite{supp}.

The FFTFIM, with $h_d=0$ and $h_d>0$, can be implemented on current D-Wave quantum annealing devices. While the frustrated AFM triangular AF 
Ising model had already been studied in depth \cite{king_nature, king_21}, only recently was the FFTFIM implemented on such a device \cite{ali_24}. In this recent study, only the spin order parameter was investigated, and it would be interesting to study dimer order parameter as well, in the light of our results.
\vskip2mm

\begin{acknowledgments}
{\it Acknowledgments.}---We would like to thank Fabien Alet, Ling Wang, and Sumner Hearth for useful discussions.
This work was supported by the Simons Foundation under Grant No.~511064, by the National Natural Science Foundation of China under Grants 
No.~12122502 and No.~12175015, and by the Swiss National Science Foundation under Grant No. 182179. Most of the numerical calculations were carried out on the Shared Computing Cluster managed by Boston University's 
Research Computing Services.
\end{acknowledgments}

\vspace{-6mm}

\let\oldaddcontentsline\addcontentsline
\renewcommand{\addcontentsline}[3]{}
\bibliography{fftfim.bib}

\let\addcontentsline\oldaddcontentsline

\onecolumngrid

\newpage

\begin{center}

{\bf\large Supplementary Materials for ``Primary and Secondary Order Parameters in the Fully Frustrated Transverse Field Ising Model on the Square Lattice''}

\vskip3mm

Gabe Schumm, Hui Shao, Wenan Guo, Fr\'ed\'eric  Mila, Anders W. Sandvik

\end{center}

\vskip5mm

\twocolumngrid

\setcounter{equation}{0}
\setcounter{figure}{0}
\setcounter{section}{0}
\setcounter{table}{0}
\setcounter{page}{1}

\renewcommand{\theequation}{S\arabic{equation}}
\renewcommand{\thefigure}{S\arabic{figure}}
\renewcommand{\thetable}{S\arabic{table}}
\renewcommand{\thesection}{S\Roman{section}}
\renewcommand{\thepage}{S\arabic{page}}

\tableofcontents

\section{Phase Diagram}\label{sec_1}

The finite temperature phase diagram for the FFTFIM is shown in Fig.~\ref{phase_diagram}. At temperatures above $T_{\rm c2}(\Gamma)$, or transverse field
values above $\Gamma_c$, the system is in the paramagnetic (PM) phase, while for $\Gamma < \Gamma_c$ and $T_{\rm  c1}(\Gamma) < T < T_{\rm  c2}(\Gamma)$, 
it is in a critical KT phase. For $\Gamma < \Gamma_c$ and $T < T_{\rm  c1}(\Gamma)$, the system is ordered, exhibiting $Z_8$ and $Z_4$ symmetry breaking 
in the primary and secondary order parameters, respectively. The phase diagram was constructed in the following way:

\begin{figure}[b]
\centering
\includegraphics[width=80mm]{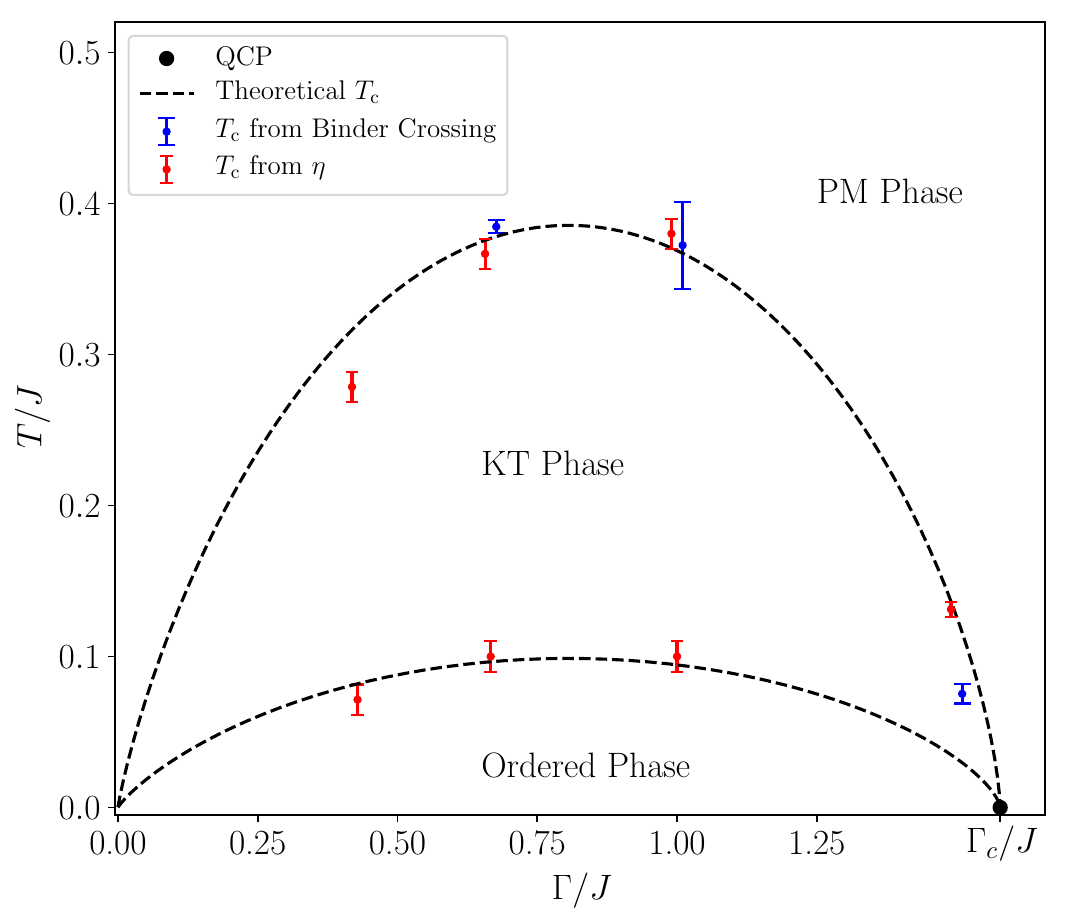}
\caption{Phase diagram of the square-lattice FFTFIM in the plane of transverse field $\Gamma$ and temperature $T$. The QCP is shown as the
black circle at $T=0$. The black dashed lines are the theoretical predictions
for the phase boundaries made in Ref.~\cite{emig, korshunov}, and the red and blue points are the transition temperatures identified using our QMC
simulations. The theoretical phase boundaries have overall unknown factors that we have adjusted for best agreement at the maximums.}
\label{phase_diagram}
\end{figure}

\begin{figure}[t]
\begin{center}
\includegraphics[width=\columnwidth]{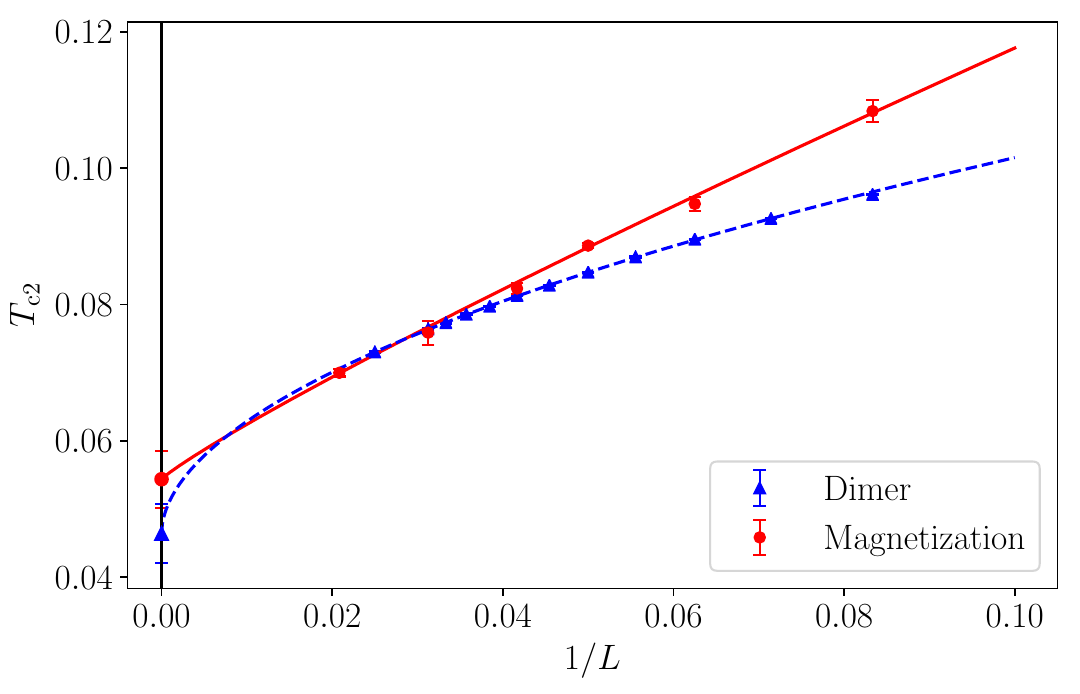}
\caption{Finite-size scaling of the Binder crossing temperature at $\Gamma/J = 1.50$. The temperatures for the magnetization (red) and dimer (blue)
order parameters were extracted for pairs of systems of size $(L, 2L)$. The error bars on the extrapolated $T_{\rm c2}(\infty)$ values were estimated
by a bootstrapping procedure.}
\label{KT_FSS}
\end{center}
\end{figure}

To locate the upper transition temperature $T_{\rm c2}$, we define Binder cumulants for both order parameters \cite{binder_1}:
\begin{equation}  \label{binder_eq}
 U_d  = 2 - \frac{\langle |\Psi|^4\rangle}{\langle |\Psi|^2\rangle^2}, \quad 
 U_m  = 2 - \frac{\langle |m|^4\rangle}{\langle |m|^2\rangle^2}.
\end{equation}
At the critical temperature, the Binder cumulant should be system-size dependent, and we indeed find that curves of $U$ versus $T$ for different $L$
values intersect each other close to a point $T_{\rm c2}$. We extract the temperatures at which the cumulants for pairs of system
sizes $L$ and $2L$ intersect, and extrapolate these to the infinite-size $T_{\rm c2}$ using the expected finite-size scaling form (shift of a definition of the critical point with the system size) at a KT transition \cite{hsieh_2013}:
\begin{equation} \label{binder_fss_eq}
\begin{split}
T_{c2}(L) = T_{c2}(\infty) + \frac{a}{\log^2(aL)},
\end{split}
\end{equation}
An example of this procedure is shown in Fig.~\ref{KT_FSS} for $\Gamma/J = 1.50$. The fits to the primary (red) and secondary (blue) binder crossing
temperatures give infinite-size $T_{c2}$ valuess that agree well with each other to within error bars estimated using bootstrapping of the data.

At the lower phase boundary, between the KT phase and the ordered phase, we do not find any crossings of the Binder cumulants. Instead, we located
$T_{\rm c1}$ from the expected power-law scaling of the magnetic and dimer order parameters, with exponents $\eta_m=1/16$ and $\eta_d=1/4$ at the lower
boundary, supported also by the observed  break-down of power-law scaling of the order parameters below $T_{\rm c1}$. Our analysis of the correlation
functions is described in the main text and in more detail in Sec.~\ref{sec_2}.

The black point at $T=0$ in Fig.~\ref{phase_diagram} is the quantum critical point (QCP), which was identified in Ref.~\cite{wenzel_mila}; we further
refined its location in the present work as detailed below in Sec.~\ref{sec_5}. The black dashed lines show the theoretical scaling
of $T_{\mathrm{c}i}$ versus $\Gamma$ predicted by Ref. \cite{emig,korshunov}, in which an overall factor has been
adjusted to fit our data at the maximums. These phase boundaries are formally only valid for $\Gamma_c - \Gamma \ll \Gamma_c$ (near the QCP), but we
include them here all the way to $\Gamma=0$ as a visual aid.

\section{Height Model Mapping}\label{sec_2}

Here, we make explicit the mapping between the spins, dimers, and height variables of the FFTFIM. Following the prescription in Ref.~\cite{emig,korshunov},
we start by identifying the 2-to-1 mapping between the ground states of the classical fully frustrated Ising model and those of the purely kinetic dimer
model on the square lattice \cite{villain_dimer}. These degenerate states are defined by the configurations where each plaquette contains a single frustrated
bond. The dimer model on the square lattice has been studied extensively \cite{fisher_dimer,rokhsar_kivelson_dimer,mila_dimer} and its height representation
is known \cite{henley_97}. We repeat the mapping here for the specific sake of clarifying the relationship between the spin and dimer order parameters,
which was not emphasized in previous treatments.

\begin{figure}[t]
\centering
\includegraphics[width=60mm]{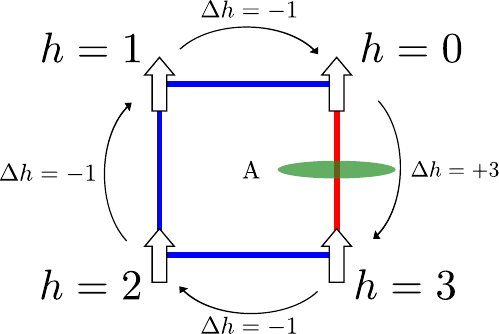}
\caption{Height model mapping rules for a plaquette at dual sublattice $A$. Height variables $h$ are labeled at the spin sites.}
\label{height}
\end{figure}

The mapping proceeds as follows: begin by dividing the dual lattice into two sublattices $A$ and $B$, where dimers connect dual lattice sites from sublattice
$A$ to $B$. Assign an arbitrary height value to a site on the spin lattice and move clockwise around a dual lattice site. If the dual lattice site is on
sublattice $A$ and a dimer is crossed, the height value at the following spin lattice site is increased by 3, otherwise it is reduced by 1; see
Fig.~\ref{height}. If the dual lattice site is on sublattice $B$, the same rules apply with signs reversed; -3 and +1, respectively.

In the ordered phase, the dimers are ordered along columns or rows, corresponding to a height profile that is "flat," i.e., the height value at any
given lattice site is bounded from above. In the critical and disordered phases, plaquettes with three frustrated bonds begin to populate the system.
The height model is then "rough", with a logarithmically diverging height profile. This behavior can be described  by an effective elastic free energy
with a periodic "locking" potential favoring flat height configurations \cite{burton_97,mila_dimer};
\begin{equation}\label{free_energy}
F(\{h(\vec{r})\}) = \int d\vec{x} \,\left[\frac{K}{2} \big|\vec{\nabla} h(\vec{r})\big| + V\cos\left(2\pi h(\vec{r})\right) \right],
\end{equation}
where $h(\vec{r})$ is the coarse-grained height field (height variables averaged over a plaquette), $K$ is the temperature dependent stiffness constant,
and $V>0$ is the strength of the locking potential. 

As discussed in the main text, for  $T<T_{c1}$ the locking potential is a relevant perturbation and the system orders \cite{jose,henley_97}.
For $T_{c1}<T<T_{c2}$, "screw dislocations" begin to emerge in pairs throughout the system. These defects are the aforementioned plaquettes containing
not one but three frustrated bonds, which appear bound in neutral pairs. For $T>T_{c2}$, the interaction between the defects becomes sufficiently weak
for the pairs to unbind, spreading throughout the system and disordering the critical phase \cite{emig,emig_06,korshunov}. 

\begin{figure}[t]
\centering
\includegraphics[width=80mm]{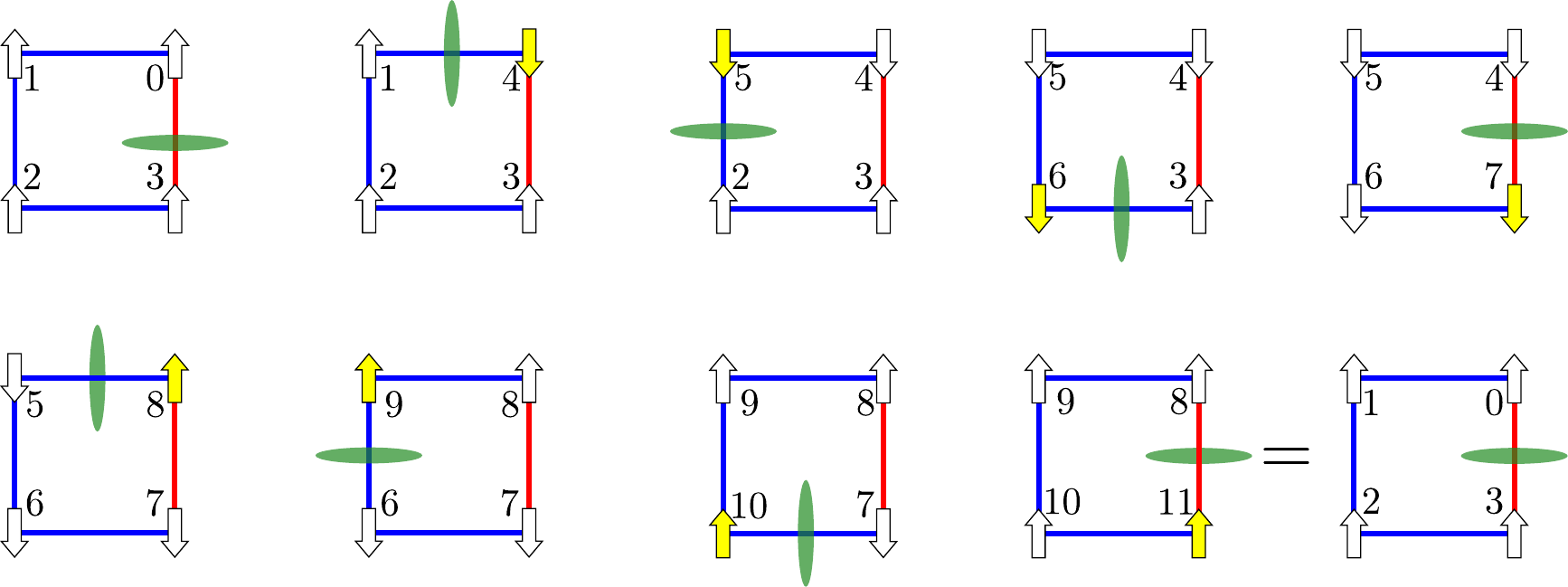}
\caption{Demonstration of the periodicity of the spin variables with the height variables. Upon traversing the plaquette, each spin (marked in yellow) 
is flipped sequentially (left to right, upper then lower row) and the height variables are computed according to the rules laid out in the text. The 
final spin configuration is identical to the initial one, but with each height increased by 8, implying that the spins must be periodic in the heights 
with period 8.}
\label{period}
\end{figure}

In the critical (rough) phase, the locking potential is irrelevant, and the system is described by a free Gaussian theory. The long distance height
difference correlations can be easily calculated \cite{blote_84, henley_97, jose}:
\begin{equation} \label{height_corr}
\begin{split}
	\frac{1}{2} \langle \big| h(\vec{r}) - h(\vec{0}) \big| ^2\rangle \propto \frac{ \ln(r)}{2\pi K}.
	\end{split}
\end{equation}
We relate this form to the spin-spin and dimer-dimers correlations by noting that the spins are periodic in the height variables, with a period of 8.
This can be seen by turning the spins around a plaquette twice as depicted in Fig.~\ref{period}. 

This periodicity implies that the spins can be written as a Fourier series:
\begin{equation} \label{fourier}
\begin{split}
	\sigma^z(\vec{r}) = \sum_G \tilde{\sigma}^z_G e^{iGh(\vec{r})}
\end{split}
\end{equation}
where $G= \frac{2\pi}{8} n $ with $ n = 1, 2, 3, \dots$. The spin-spin correlations for a single Fourier mode is given by 
\begin{equation} 
\begin{split}
	\langle e^{iGh(\vec{r})}e^{-iGh(\vec{0})}\rangle =  e^{-\frac{1}{2}G^2\langle | h(\vec{r})-h(\vec{0}) |^2\rangle},
\end{split}
\end{equation}
and clearly the smallest $G$ value in the expansion in Eq.~(\ref{fourier}) will dominate. Thus, we can express the coarse-grained spin and dimers as
\begin{equation} \label{spin_dimer}
\begin{split}
	&\sigma^z(\vec{r}) \propto e^{i\frac{2\pi}{8}h(\vec{r})} \\
	&d(\vec{r}) \propto \sigma^z(\vec{x}) \sigma^z(\vec{r} +\hat{x})   \approx \left(\sigma^z(\vec{r})\right)^2 \propto e^{i\frac{4\pi}{8}h(\vec{r})}.
	\end{split}
\end{equation}

Plugging in the height difference correlations in the rough phase, Eq.~(\ref{height_corr}), we can extract the anomalous dimensions for the spins and dimers:
\begin{equation} \label{spin_dimer_corr}
\begin{split}
	 \langle\sigma^z(\vec{x})\sigma^z(\vec{0}) \rangle &\propto  e^{-\frac{1}{2}\left(\frac{2\pi}{8}\right)^2\langle | h(\vec{r})-h(\vec{0}) |^2\rangle} \\
	& \propto r^{-{\pi}/{32K}} \propto r^{-\eta_m}\\
 \langle d(\vec{r})d(\vec{0}) \rangle &\propto  e^{-\frac{1}{2}\left(\frac{24pi}{8}\right)^2\langle | h(\vec{r})-h(\vec{0}) |^2\rangle} \\
	&\propto r^{-{\pi}/{8K}} \propto r^{-\eta_d}	\end{split}
\end{equation}
This mapping implies that $\eta_d = 4 \eta_m$, which we have confirmed numerically in Fig.~3c in the main text.

The above derived relationship between $\eta$ and $K$ allows us to determine the values of anomalous dimension at the boundaries of the critical phase.
Because the effective free energy in Eq.~(\ref{free_energy}) is exactly that of the 2D classical XY model with an eight-state clock anisotropy term, we
can use the RG analysis preformed in Ref.~\cite{jose} to determine the critical values of the coupling $K$ separating the critical phase from the
ordered and disordered phases.

The lower and upper critical temperatures correspond to $K_{c1} = \pi/2$ and  $K_{c2} = \pi/16$, between which the system is critical
(cf.~with Ref.~\cite{blote_84}). The values of $\eta_m$ at the phase boundaries are then
\begin{equation}
\eta_m(T_{c1}) = 1/16,~~~
\eta_m(T_{c2}) = 1/4.
\end{equation}

\section{Reduced $\chi^2$ for Power Law Fit in Critical Phase}\label{sec_3}

\begin{figure}[t]
\begin{center}
\includegraphics[width=80mm]{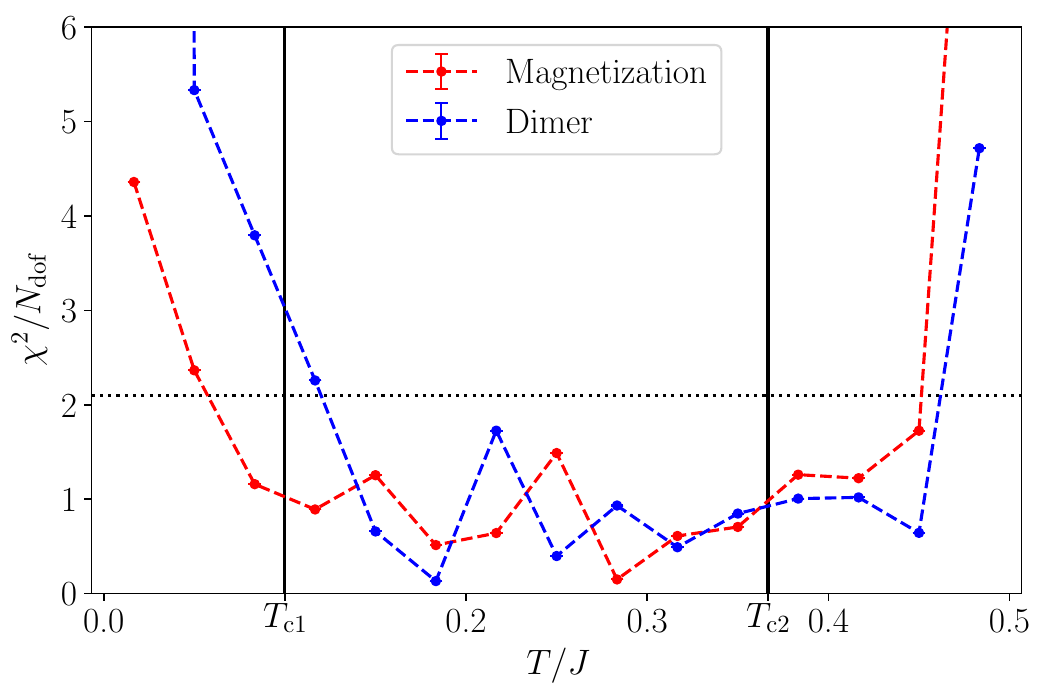}
\caption{Reduced $\chi^2$ values from power law fits used to extract $\eta_m$ (red) and $\eta_d$ (blue) in Fig.~3 of the main text. The black dashed
line denotes the value corresponding to the 95\% confidence level for the number of degrees of freedom of the fits. The reduced $\chi^2$ begins
to exceed the 95\% confidence level below $T_{c1}$ and above $T_{c2}$, indicating that the order parameters no longer scale as power laws.}
\label{chi}
\end{center}
\end{figure}

Below the lower critical temperature $T_{c1}$, the order parameters should no longer scale as power laws. Around this temperature, we therefore expect
that the reduced $\chi^2$ from a power law fit used to extract $\eta_m$ and $\eta_d$ would begin to degrade. Figure \ref{chi} shows the reduced $\chi^2$
for these fits versus temperature compared to the reference value corresponding to the 95\% confidence level (black dashed line). The results are indeed
consistent with the power law form no longer being applicable below $T_{c1}$ and above $T_{c2}$. Larger system sizes would be required to observe the
fits deteriorating more significantly close to the phase boundary. The results shown here were obtained by including the largest 8 system sizes in
Figs.~3(a) and 3(b) in the fits.

\section{Correlation Function at Zero Temperature }\label{sec_4}

At the quantum critical point, we measure the long distance correlation functions to extract the scaling exponents $\eta_m$ and $\eta_d$ at $T=0$.
The spin-spin correlation function is defined as 
\begin{equation}\label{CFT_CM}
\lim_{r \to \infty} C_M(\vec{r}) = \lim_{r\to \infty} \langle \sigma^z_{m_i}(\vec{r}) \sigma^z_{m_i}(0) \rangle \sim \frac{1}{r^{1+\eta_m}},
\end{equation}
where $\sigma^z_{m_i}(\vec{r})$ is the spin operator at a site $\vec{r}$ on sublattice $i$ and we expect $\eta_m = \eta_{\mathrm{3D}XY}$. In connection to the relevant field theory,
$1+\eta_{\mathrm{3D}XY}=2\Delta_\phi$, where $\Delta_\phi \approx 0.519088$ \cite{OPE_O2_1} is the scaling dimension of the order parameter
of the 3D $O(2)$ model. This reflects the known irrelevance of $Z_q$ perturbations with $q\ge 4$ at this critical point.

The dimer-dimer correlation function is defined as
\begin{equation}\label{CFT_CD}
D_x(\vec{r}) =\langle d_{x}(\vec{r}) d_{x}(0) \rangle,
\end{equation}
where $d_{x}(\vec{r})$ is the $x$-oriented dimer operator at a site $\vec{r}$. However, in the ordered phase, the dimer correlations will oscillate
between smaller and larger values and the order parameter of interest is the difference between these modulated correlations;
\begin{equation}
D^*_x(\vec{r}) = D_x(\vec{r}) -\frac{D_x(\vec{r} - \hat{y}) + D_x(\vec{r}+\hat{y})}{2}.
\end{equation}
Averaging over both horizontal and vertical dimer orientations, the expected scaling is 
\begin{equation}
\lim_{r \to \infty} C_D(\vec{r}) = \lim_{r\to \infty} \frac{D^*_x(\vec{r}) +D^*_y(\vec{r})}{2} \sim \frac{1}{r^{1+\eta_d}}
\end{equation}
We evaluate both of these correlation functions at the largest separation for a periodic, finite system of linear size $L$, $\vec{r} = (L/2, L/2)$. 
In Fig. 4 of the main text we show that the dimer correlations fall off with distance as a power law with exponent corresponding to the leading charge-2 operator
of the 3D $O(2)$ model; $1+\eta_d = 2\Delta_t$, with  $\Delta_t \approx  1.23629$ \cite{OPE_O2_1}.

\section{Determining $\mathbf{\Gamma_c}$}\label{sec_5}

To determine $\Gamma_c$, we located the value of $\Gamma$ where the primary order parameter correlation functions scales as power law with the
expected power $\Delta_\phi$. We measured the correlations function for $\Gamma$ values in a small range around the previously reported $\Gamma_c$
and compared the $\chi^2$ from power law fits to the data at these different values. To reduce the error bars on the measured data, we interpolated
the correlation function data for a given system size using second-order polynomials for several data sets in the $\Gamma$ range $\left[1.576, 1.578\right]$. 

\begin{figure}[t]
\begin{center}
\includegraphics[width=\columnwidth]{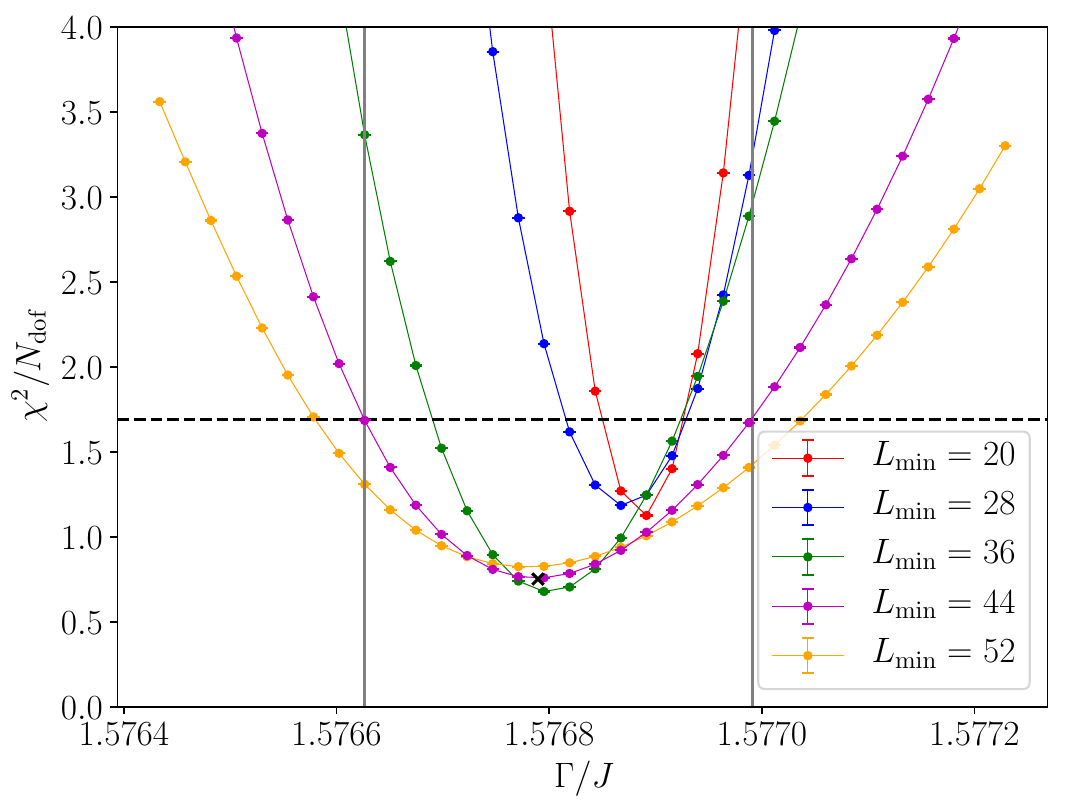}
\caption{Reduced $\chi^2$ from power-law fits to the spin-spin correlation function data versus $\Gamma$. The $\chi^2$ minimum converges as the
smallest $L$ value used in the fit is increased. Using the number of degrees of freedom for the $L_{\rm min} = 44$ fit, we determine the $\Gamma$ values
within the 95\% confidence interval of the $\chi^2$ distribution, indicated here by the two vertical black lines. The error bars were computed by
bootstrapping the raw binned data sets.}
\label{cm_chi}
\end{center}
\end{figure}

Reduced $\chi^2$ versus $\Gamma$ for the spin-spin correlation function is shown in Fig.~\ref{cm_chi}, where we see convergence in the location of
the $\chi^2$ minimum as the minimum system size used for the fit is increased. We judge that any remaining systematic errors from scaling corrections
are negligible for $L_{\rm min} = 44$, and the resulting critical point is then $\Gamma_c = 1.57680 \pm 0.00009$. To determine the error bar on $\Gamma_c$,
we located $\Gamma$ values where the reduced $\chi^2$ is equal to the the 95\% confidence level for the number of degrees of freedom used, denoted by the two vertical black lines.

\section{Secondary Order Parameter in 3D Classical Clock Models}\label{sec_6}

The 3D classical $q$-state clock model is defined by the Hamiltonian
\begin{equation}
H_q = -\sum_{\langle i, j \rangle} \cos\left(\theta_i - \theta_j\right) - h_q\sum_i \cos\left(q\theta_i\right),
\end{equation}
with $\theta_i \in \left [ 0, 2\pi \right)$ defining the orientation of a two-component spin on cubic lattice site $i$. At fixed $h_q$,
below $T=T_{\rm c}$, the systems orders into a $Z_q$ symmetry breaking phase. The $Z_q$ perturbation for $q\ge 4$ is known to be irrelevant 
at the critical point, so that the 3D $O(2)$ universality class applies (though changing the symmetry broken in the ordered phase).

For $q=8$, the microscopic symmetries of the model exactly match those of the square-lattice FFTFIM, and we here study this clock model as
a bench-mark case of primary and secondary order parameters. Just as in the FFTFIM, we can define charge-1 and charge-2 scalar order parameters
for the clock models:
\begin{equation}
\begin{split}
m_x = \sum_i \cos(l\theta_i)&,~~m_y = \sum_i \sin(l\theta_i),\\
m_l^2 = &\langle m_x ^2\rangle  + \langle m_y ^2\rangle,
\end{split}
\end{equation}
where $l$ labels the charge.

\begin{figure}[b]
\begin{center}
\includegraphics[width=70mm]{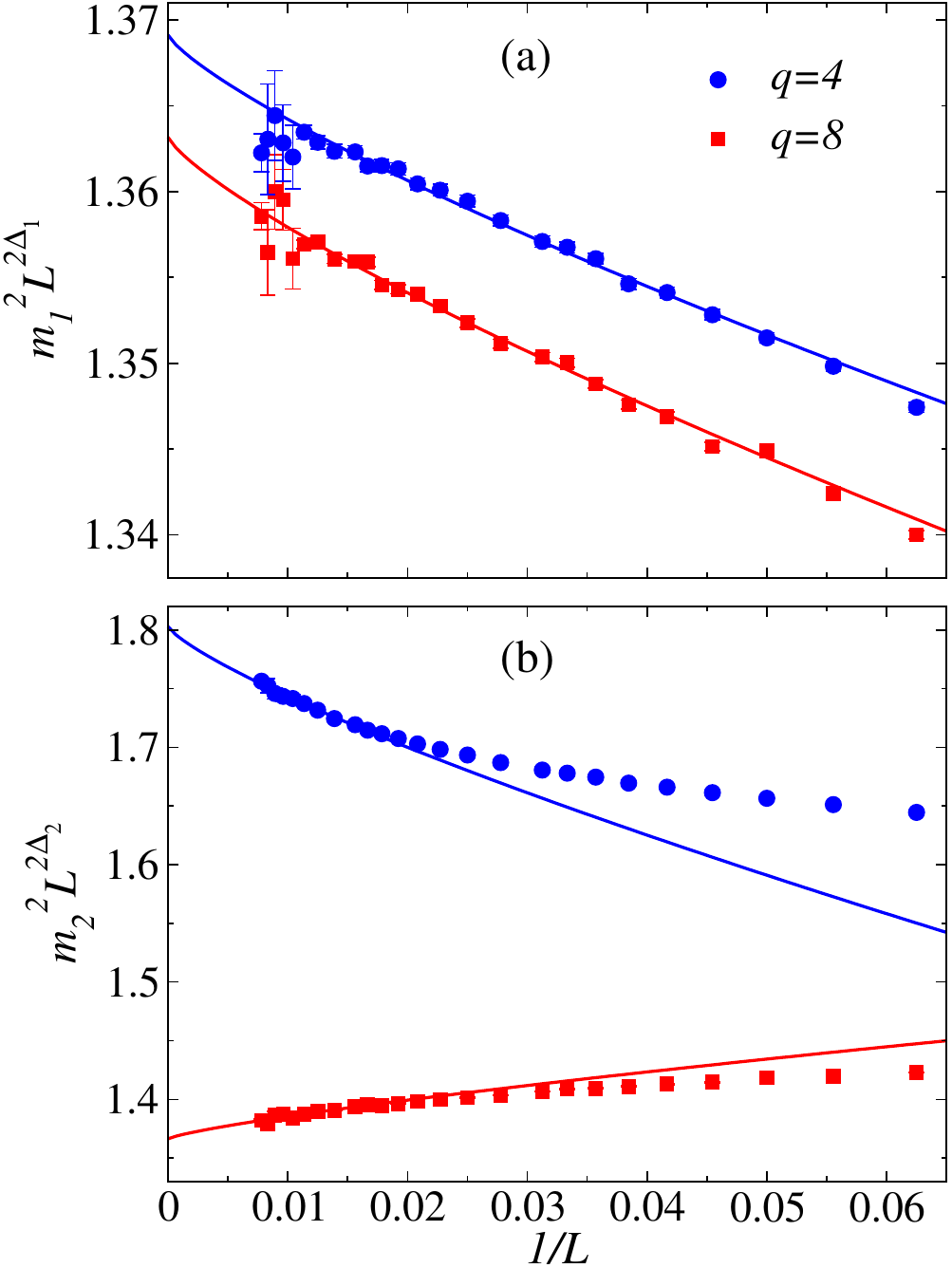}
\caption{Scaling behaviors of $m_l^2$, with charge $l=1$ i (a) and $l=2$ in (b), at the estimated critical temperatures for the $q=4$ and $q=8$ 
clock models (using $1/T_c=0.45416467$ for $q=8$ \cite{hasenbusch_19} and $T_c=2.20467$ for $q=4$. The curves show the form Eq.~(\ref{mlforms}) 
with only the constants $a$ and $b$ adjusted for best fits and with results for the smaller system sizes exccluded systematically until a
statistically good reduced $\chi^2$ value is obtained.}
\label{clock_crit_scaling}
\end{center}
\end{figure}

Although our primary interest here is in the $q=8$ case, we carried out simulations at the critical points of three different versions of the model,
namely, the XY model (with no clock field, corresponding to $q\to \infty$), the 8-state hard clock model (with the spins restricted to the $q$ discrete 
states, corresponding to $h_8 \to \infty$), and the 4-state soft clock model with anisotropic field $h_4 = 1$. We expect the same scaling dimensions in all 
cases on account of the irrelevance of the clock perturbations. The critical points of the first two models can be found in Ref.~\cite{hasenbusch_19}, 
while $T_c=2.20465(1)$ of the third model was reported in \cite{shao_2020}. In an improved analysis, we found this value to be about two error bars too 
low and here report results obtained with $T=2.20467$.

\begin{figure}[t]
\begin{center}
  \includegraphics[width=70mm]{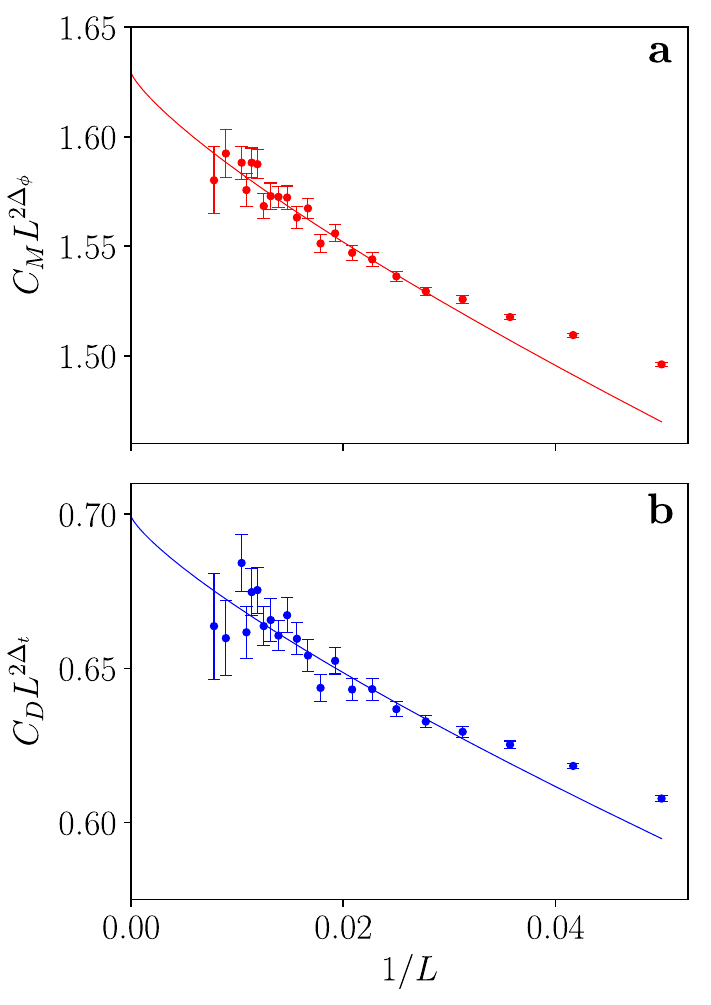}
\caption{Scaling behaviors of $C_M(L/2)$ and  $C_D(L/2)$, plotted as in Fig. \ref{clock_crit_scaling}.}
\label{FFTFIM_crit_scaling}
\end{center}
\end{figure}

In Fig.~\ref{clock_crit_scaling} we fit our data for $q=4$ and $q=8$ to the expected scaling form with the leading finite-size correction included;
\begin{equation}
m_l^2 L^{2\Delta_l} = a + bL^{-\omega}
\label{mlforms}
\end{equation}
with $\Delta_1 = \Delta_\phi = 0.519088$, $\Delta_2 = \Delta_t = 1.23629$ \cite{OPE_O2_1}, and $\omega = 0.789$ \cite{hasenbusch_19}. Only the prefactors
$a$ and $b$ were fitted. In the case of the XY model, the results fall so close to those for $q=8$ that we have not included them in the figure
for clarity (we also note that the error bars are larger for the XY model after simulations of comparable length).

To compare the clock results
more directly to the FFTFIM results shown in Fig.~4 of the main text, in Fig.~\ref{FFTFIM_crit_scaling} we show the FFTIFM data analyzed
in the same way as the clock results in Fig.~\ref{clock_crit_scaling}, with the expected leading power laws of $L$ divided out and fitting only to
the expected correction given by Eq.~(\ref{mlforms}). Though the sign of the (non-universal) amplitude of the correction to the dimer correlations
in Fig.~\ref{FFTFIM_crit_scaling}(b) is different from that in the $q=8$ result in Fig. \ref{clock_crit_scaling}(b), the overall behavior of the
corrections (the leading term fitted to and the higher-order corrections causing deviations from the fit) is similar. The primary order
parameter of the FFTFIM in Fig.~\ref{FFTFIM_crit_scaling}(a) seems to have more significant subleading corrections than the clock results in
Fig.~\ref{clock_crit_scaling}, but the large-$L$ asymptotic form is again similar and consistent with the leading correction.

\begin{figure}[b]
\begin{center}
\includegraphics[width=83mm]{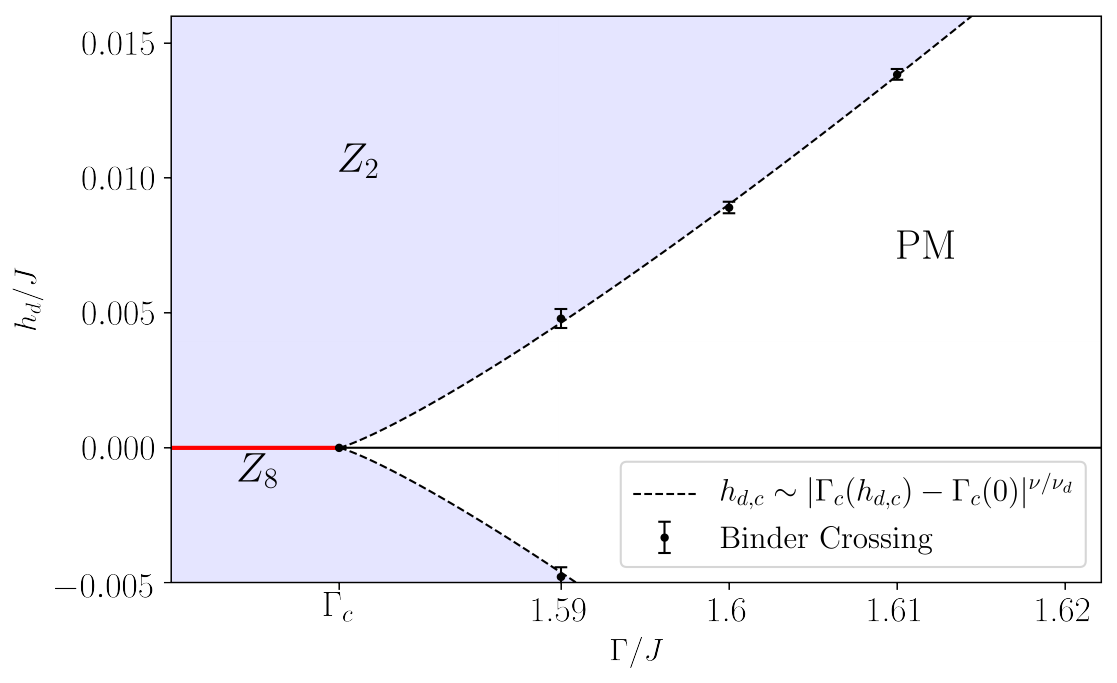}
\caption{$\Gamma$--$h_d$ phase boundary. The red line indicates the $Z_8$-breaking phase that exists only at $h_d=0$.
The black dashed lines show the boundaries between the paramagnetic and $Z_2$-breaking phases and  are of the form given in
Eq.~(\ref{phase_bound_form}), where a single prefactor was adjusted to fit the data.}
\label{Z2_phase_bound}
\end{center}
\end{figure}

\section{$\mathbf{Z_2}$ Phase}

By alternating the Ising couplings betweek $J-h_d$ and $J+h_d$ ($J=1$) on either rows or columns in the FFTFIM, a single dimer state and its accompanying 
two spin states are energetically favored. Thus, this perturbation, arbitrarily weak, induces a $Z_2$ symmetry breaking phase in the plane of $\Gamma$ 
and the strength of the coupling modulation $h_d$. 

We can derive the expected asymptotic form of the phase boundary by considering the diverging length scales corresponding to each perturbing field 
($\Gamma $ and $h_d$). Upon tuning $\Gamma \to \Gamma_c$ and $h_d \to h_{d, c}$, their respective correlation lengths will diverge
as \begin{equation}
\xi_\Gamma \sim |\Gamma-\Gamma_c|^{-\nu},~~\xi_{h_d} \sim |h_d-h_{d, c}|^{-\nu_d},
\end{equation}
where $\nu$ again is the 3D $O(2)$ correlation length exponent, $\nu=(3-\Delta_s)^{-1}$ in terms of the relevant charge-0 (singlet) scaling
dimension $\Delta_s \approx 1.51136$ \cite{OPE_O2_1}, and $\nu_d=(3-\Delta_t)^{-1}$ with $\Delta_t = 1.23629$. Assuming that the $\Gamma$--$h_d$
phase boundary corresponds to a fixed ratio of these two length scales, we extract the form
\begin{equation}\label{phase_bound_form}
h_{d,c} \sim |\Gamma_c(h_{d,c}) - \Gamma_c(0)|^{\nu/\nu_d}.
\end{equation}
We have confirmed this form using numerics, as shown in Fig.~\ref{Z2_phase_bound}, using the same type of QMC simulations and analysis
as explained in previous sections.
\vfill

\end{document}